\documentclass[
  aps,
  prl,
  reprint,
  superscriptaddress,
  amsmath,
  amssymb,
  raggedbottom
]{revtex4-2}

\usepackage{graphicx}
\usepackage{dcolumn}
\usepackage{bm}
\usepackage{booktabs,tabularx}
\usepackage{xcolor}
\usepackage[
  colorlinks=true,
  linkcolor=blue,
  citecolor=blue,
  urlcolor=blue
]{hyperref}

\newcommand{\lsec}[1]{\textit{#1.}---}

\newcounter{superequation}
\makeatletter
\@addtoreset{equation}{superequation}
\makeatother

\begin{document}

\title{Liouvillian Non-Invertible Symmetry}

\author{Ying-Cheng Yang}
\affiliation{Department of Physics, National Cheng Kung University, Tainan 701401, Taiwan}
\affiliation{Physics Division, National Center for Theoretical Sciences, Taipei 106319, Taiwan}

\author{Jhen-Dong Lin}
\email{jhendonglin@gmail.com}
\affiliation{Department of Applied Science, National Taitung University, Taitung 950, Taiwan}

\author{Yueh-Nan Chen}
\email{yuehnan@mail.ncku.edu.tw}
\affiliation{Department of Physics, National Cheng Kung University, Tainan 701401, Taiwan}
\affiliation{Physics Division, National Center for Theoretical Sciences, Taipei 106319, Taiwan}
\affiliation{Center for Quantum Frontiers of Research and Technology (QFort), National Cheng Kung University, Tainan 701401, Taiwan}

\begin{abstract}
Non-invertible symmetries extend conventional symmetries beyond group-like transformations and play an important role in quantum field theory and many-body physics. However, a general framework of non-invertible symmetries in open quantum systems is still lacking. Here, we formulate a general Liouville-space duality framework that extends Hilbert-space non-invertible dualities to open quantum systems. The resulting defect relates Liouvillians with dual Hamiltonian and dissipative parameters and becomes an exact symmetry at self-duality. We apply this framework to the periodic transverse-field Ising chain with two collective Lindblad channels exchanged by Kramers–Wannier (KW) duality. A strong $\mathbb Z_2$ symmetry divides Liouville space into four parity sectors, while the defect acts nontrivially only in the even-even sector. At self-duality, the KW defect imposes a momentum-resolved selection rule on Liouvillian Jordan blocks. Modes with different Liouville momenta or KW-defect eigenvalues cannot form a common exceptional Jordan block. For a three-site chain, a global Rényi-2 KW asymmetry dynamically identifies the self-dual conditions. At the same time, the spectral structure exhibits both a semisimple degeneracy and a self-dual exceptional point with a perturbation-selective response. These results extend non-invertible dualities to Markovian quantum dynamics and establish a symmetry-based selection rule for Liouvillian Jordan structures.
\end{abstract}

\maketitle

\lsec{Introduction}
Symmetries in quantum many-body systems are conventionally represented by invertible operators that obey group composition laws. They organize phases and spectra through symmetry sectors, conservation laws, and selection rules~\cite{McGreevy2023,Gaiotto2015}. Non-invertible symmetries broaden this framework by admitting symmetry operators without inverses and more general fusion rules~\cite{Shao2023,SchaferNameki2024,JiWen2020}. They have been identified in quantum field theory~\cite{Chang2019,Choi2022,Kaidi2022}, statistical mechanics~\cite{Aasen2016,JacobsenSaleur2023}, and quantum lattice models~\cite{Lootens2023,SeifnashriShao2024,Cao2024,LiOshikawaZheng2023,Chatterjee2024,EckFendley2024,Chung2026}, where they impose constraints beyond those of ordinary group symmetries~\cite{Seiberg2024,LiMitra2026,Apte2023}. A paradigmatic lattice realization is Kramers--Wannier (KW) duality in the transverse-field Ising model (TFIM)~\cite{KramersWannier1941,SeibergShao2024,Frohlich2004,ZhangSierra2025}. On a periodic chain, the non-invertible KW operator combines the duality transformation with a projection onto the even parity sector~\cite{Shao2023,Seiberg2024}.

Most lattice constructions of non-invertible symmetry have been developed for closed quantum systems. Recent studies have considered related ideas in quantum operations~\cite{OkadaTachikawa2024} and mixed states~\cite{Sun2025,SchaferNamekiMixed2025}. However, a direct formulation for a Markovian Liouvillian has not yet been developed. In Lindblad dynamics~\cite{Gorini1976,Lindblad1976}, the Liouvillian acts on operator space. Strong symmetries can act independently on the ket and bra indices, giving a finer sector structure than in Hilbert space~\cite{Buca2012,Albert2014}. This raises a natural question: how should a Hilbert-space non-invertible duality be lifted to Liouville space when the duality acts simultaneously on Hamiltonian couplings and dissipative parameters?

Two complementary aspects of the resulting Liouvillian structure are particularly relevant. Liouvillian spectra can host exceptional points (EPs), where eigenvalues and eigenmodes coalesce~\cite{Heiss2012,Bergholtz2021,Minganti2019,arkhipov2020liouvillian,hu2025asymptotic}, giving rise to characteristic dynamical~\cite{Chen2022,SunYi2024} and perturbative responses~\cite{Wiersig2022}. This raises the question of whether non-invertible symmetry constrains the Jordan structure associated with exceptional degeneracies. A complementary dynamical probe is provided by entanglement asymmetry~\cite{ma2022symmetric,vaccaro2008tradeoff,gour2009measuring,Ares2023}, which has also been studied in dissipative dynamics~\cite{Caceffo2024}. Its extensions to non-invertible symmetries~\cite{Benini2025} and specifically to KW duality~\cite{Vescovo2026} motivate a corresponding diagnostic of self-duality in Markovian dynamics. The dissipative TFIM therefore provides a setting in which both symmetry constraints on exceptional degeneracies and dynamical signatures of self-duality can be investigated.

In this work, we develop a general Liouville-space framework for non-invertible symmetry in Markovian dynamics. We realize this construction in the periodic TFIM with two collective Lindblad channels exchanged by KW duality. The duality relates coherent and dissipative parameters and becomes an exact symmetry at self-duality. Strong $\mathbb Z_2$ symmetry divides Liouville space into four parity sectors, with nontrivial KW-defect action confined to the even--even sector. We use a global R\'enyi-2 KW asymmetry to dynamically identify the self-dual conditions. We then derive a selection rule for Liouvillian Jordan blocks by combining translation symmetry with the KW defect. The $N=3$ spectrum illustrates this rule through exceptional and semisimple degeneracies. Finally, we examine the first-order response of a self-dual EP2 to local perturbations.

\lsec{Liouvillian non-invertible symmetry}
We begin with the non-invertible symmetry condition in Liouville space. Let $D$ be a non-invertible operator acting on the Hilbert space. Its action on operators is $\Phi_D(\rho)=D\rho D^\dagger$. Under column vectorization, $|\rho\rangle\!\rangle=\operatorname{vec}(\rho)$, this map is represented by $\mathbb D=D^*\otimes D$. An exact Liouvillian non-invertible symmetry satisfies
\begin{equation}
[\mathbb L,\mathbb D]=0.
\label{eq:Liouvillian-NIS}
\end{equation}

More generally, the duality may relate Liouvillians at different parameter points. Consider a Markovian Liouvillian $\mathcal L_{\boldsymbol\lambda}$, where $\boldsymbol\lambda$ includes both coherent and dissipative parameters,
\begin{equation}
\dot{\rho}=\mathcal L_{\boldsymbol\lambda}(\rho)=-i[H_{\boldsymbol\lambda},\rho]+\sum_\mu \mathcal C[L_{\mu,\boldsymbol\lambda}](\rho),
\label{eq:general_GKSL}
\end{equation}
with $\mathcal C[L](\rho)=L\rho L^\dagger-\frac{1}{2}\{L^\dagger L,\rho\}$. If the duality maps $\boldsymbol\lambda\mapsto\sigma(\boldsymbol\lambda)$, its Liouville space extension satisfies the duality relation
\begin{equation}
\mathbb D\,\mathbb L(\boldsymbol\lambda)=\mathbb L[\sigma(\boldsymbol\lambda)]\,\mathbb D.
\label{eq:general_Liouvillian_intertwining}
\end{equation}
This relation connects Liouvillians at dual parameter points and requires the coherent and dissipative contributions to transform consistently under the duality. Since $\mathbb D$ is non-invertible, it does not define a similarity transformation. At a self-dual point, $\sigma(\boldsymbol\lambda)=\boldsymbol\lambda$, Eq.~\eqref{eq:general_Liouvillian_intertwining} reduces to Eq.~\eqref{eq:Liouvillian-NIS}.

\begin{figure}[t]
\centering
\includegraphics[width=1\columnwidth]{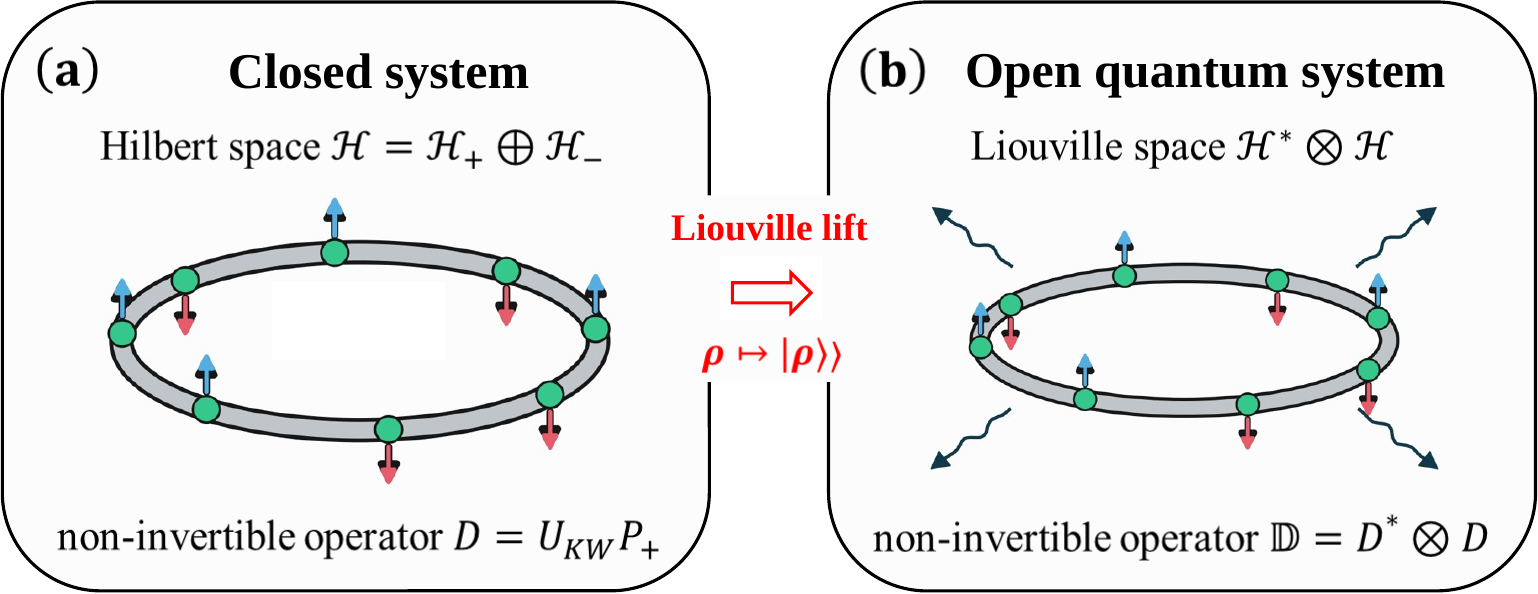}
\caption{Liouville-space extension of Kramers--Wannier non-invertible symmetry. (a) The periodic TFIM with non-invertible operator $D$. (b) The corresponding open-system construction, in which the closed-system non-invertible operator is lifted to Liouville space and the outward arrows indicate dissipation into the environment.}
\label{fig:model}
\end{figure}

We now connect this general structure to KW duality in the periodic transverse-field Ising model (TFIM), shown in Fig.~\ref{fig:model}(a). The Hamiltonian is
\begin{equation}
H=-J\sum_{j=1}^{N} Z_jZ_{j+1}-g\sum_{j=1}^{N} X_j,
\label{eq:H}
\end{equation}
where $Z_{N+1}\equiv Z_1$. Defining $A=\sum_j X_j$ and $B=\sum_j Z_jZ_{j+1}$, we have $H(J,g)=-JB-gA$. The spin flip operator $\eta=\prod_{j=1}^{N}X_j$ generates a $\mathbb Z_2$ symmetry, $[H,\eta]=0$. The Hilbert space is therefore divided into the two parity sectors $\mathcal H=\mathcal H_+\oplus\mathcal H_-$, where $\eta=\pm1$ and $P_\pm=(I\pm\eta)/2$.
Following Ref.~\cite{Shao2023}, the lattice KW operator is
\begin{equation}
D=U_{\rm KW}P_+,
\label{eq:D-def}
\end{equation}
where $U_{\rm KW}$ implements the KW transformation. We use the lattice normalization of Ref.~\cite{Shao2023} throughout~\cite{NOTE01}. The local KW relations give $DA=BD$ and $DB=AD$, and therefore $DH(J,g)=H(g,J)D$. At $J=g$, the defect becomes an exact symmetry of the Hamiltonian, $[H,D]=0$. Because $DP_-=0$, the odd parity sector is annihilated by $D$. Thus the KW defect acts nontrivially only in the even parity sector in Hilbert space.

The same projection structure becomes finer in Liouville space. An arbitrary operator can be decomposed according to the parity of its ket and bra indices,
\begin{equation}
\rho=P_+\rho P_++P_+\rho P_-+P_-\rho P_++P_-\rho P_-.
\label{eq:operator-parity-decomposition}
\end{equation}
Under vectorization, $\Pi_{pq}=P_q^*\otimes P_p$ selects the four components, with $p,q=\pm1$. The two parity sectors in Hilbert space therefore become four sectors in Liouville space. The KW defect in Liouville space selects only one of these sectors. Since $D=P_+DP_+$, its lift satisfies
\begin{equation}
\mathbb D=\Pi_{++}\mathbb D\Pi_{++}.
\label{eq:defect-support}
\end{equation}
Equivalently,
\begin{equation}
\Phi_D(\rho)=D(P_+\rho P_+)D^\dagger.
\label{eq:general_defect_support}
\end{equation}
Hence, the defect acts nontrivially only on the even--even component $P_+\rho P_+$. The $(+,-)$, $(-,+)$, and $(-,-)$ components are annihilated and belong to the kernel of $\mathbb D$. This is the Liouville-space counterpart of the projection onto the $\eta=+1$ sector in Hilbert space. The correspondence between the symmetry structures in Hilbert and Liouville spaces is summarized in Table~\ref{tab:algebra}.

The remaining question is whether the open-system dynamics also preserve these four operator sectors. We now give a concrete realization in which the same KW duality exchanges the coherent and dissipative parameters.

\begin{table}[t]
\caption{
Hilbert- and Liouville-space symmetry structure of the periodic TFIM.
}
\label{tab:algebra}
\centering
\scriptsize
\setlength{\tabcolsep}{3pt}
\renewcommand{\arraystretch}{1.05}

\begin{tabular}{
@{}
>{\raggedright\arraybackslash}p{0.47\columnwidth}
>{\raggedright\arraybackslash}p{0.47\columnwidth}
@{}
}
\hline\hline
Hilbert space & Liouville space
\\
\hline

Spin flip:
$\eta=\prod_j X_j,\ \eta^2=I$
&
$\eta_c=\eta^*\otimes\eta=\eta_L\eta_R$
\\

Translation:
$T,\ T^N=I$
&
$\mathbb T=T^*\otimes T,\ 
\mathbb T^N=\mathbb I$
\\

Projection:
$P_+=(I+\eta)/2$
&
$\Pi_{c,+}=(\mathbb I+\eta_c)/2
=\Pi_{++}+\Pi_{--}$
\\

KW operator:
$D=U_{\rm KW}P_+$
&
$\mathbb D=D^*\otimes D$
\\

Defect support:
$D^\dagger D=P_+$
&
$\mathbb D^\dagger\mathbb D
=\Pi_{++}=P_+^*\otimes P_+$
\\

$D\eta=\eta D=D$
&
$\mathbb D\eta_c=\eta_c\mathbb D=\mathbb D$
\\

KW algebra:
$D^2=P_+T$
&
$\mathbb D^2=\Pi_{++}\mathbb T$
\\

Parity sectors:
$\mathcal H=\mathcal H_+\oplus\mathcal H_-$
&
$\mathcal B(\mathcal H)
=\mathcal B_{++}\oplus\mathcal B_{+-}
\oplus\mathcal B_{-+}\oplus\mathcal B_{--}$
\\

Kernel:
$\ker D=\mathcal H_-$
&
$\ker\mathbb D
=\mathcal B_{+-}\oplus\mathcal B_{-+}\oplus\mathcal B_{--}$
\\

$DH(J,g)=H(g,J)D$
&
$\mathbb D\,\mathbb L(\boldsymbol\lambda)=\mathbb L[\sigma(\boldsymbol\lambda)]\,\mathbb D$
\\

\hline\hline
\end{tabular}
\end{table}

\lsec{Two-channel Kramers--Wannier realization}
We now realize the general construction above in the periodic TFIM by choosing two dissipative channels that are exchanged by KW duality. Since $A$ and $B$ satisfy $DA=BD$ and $DB=AD$, we introduce the two Hermitian collective jump operators
\begin{equation}
L_A=\sqrt{\gamma_A}\,A,\qquad L_B=\sqrt{\gamma_B}\,B,
\label{eq:jumps}
\end{equation}
with nonnegative rates $\gamma_A$ and $\gamma_B$. These channels act collectively on the transverse field and Ising bond operators, respectively, and the same KW transformation exchanges the two terms of the Hamiltonian. The corresponding Markovian dynamics is
\begin{equation}
\mathcal L_{J,g;\gamma_A,\gamma_B}(\rho)=-i[H(J,g),\rho]+\gamma_A\mathcal C_A(\rho)+\gamma_B\mathcal C_B(\rho),
\label{eq:two-channel-L}
\end{equation}
where $\mathcal C_O\equiv\mathcal C[O]$ for $O=A,B$, so that $\mathcal C[L_A]=\gamma_A\mathcal C_A$ and $\mathcal C[L_B]=\gamma_B\mathcal C_B$.
The KW relations exchange the two dissipative contributions. From $DA=BD$ and $DB=AD$, one obtains $\Phi_D\circ\mathcal C_A=\mathcal C_B\circ\Phi_D$ and $\Phi_D\circ\mathcal C_B=\mathcal C_A\circ\Phi_D$. Together with $DH(J,g)=H(g,J)D$, the duality acts on the full parameter set as
\begin{equation}
\sigma:(J,g,\gamma_A,\gamma_B)\longmapsto(g,J,\gamma_B,\gamma_A).
\label{eq:two-channel-duality-map}
\end{equation}
Combining these relations yields the Liouvillian KW duality relation,
\begin{equation}
\mathbb D\,\mathbb L(J,g;\gamma_A,\gamma_B)=\mathbb L(g,J;\gamma_B,\gamma_A)\,\mathbb D.
\label{eq:Liouvillian-covariance}
\end{equation}
A direct derivation from the vectorized Liouvillian is given in the End Matter. The self-dual conditions are $J=g$ and $\gamma_A=\gamma_B$. Under these conditions, Eq.~\eqref{eq:Liouvillian-covariance} reduces to $[\mathbb L,\mathbb D]=0$. The resulting Liouvillian KW non-invertible symmetry is exact for arbitrary chain length $N$.
The two-channel Liouvillian also inherits the strong $\mathbb Z_2$ symmetry. Since $H$, $A$, and $B$ commute with $\eta$, left and right multiplication by $\eta$ are separately conserved in Liouville space. With $\eta_L=I\otimes\eta$ and $\eta_R=\eta^*\otimes I$, one has $[\mathbb L,\eta_L]=[\mathbb L,\eta_R]=0$ for arbitrary $J$, $g$, $\gamma_A$, and $\gamma_B$. Defining $\Pi_{pq}=P_q^*\otimes P_p$ for $p,q=\pm1$, the four projectors resolve the independent ket and bra parities and satisfy $[\mathbb L,\Pi_{pq}]=0$. The Liouvillian therefore decomposes into four dynamically invariant blocks,
\begin{equation}
\mathbb L=\mathbb L_{++}\oplus\mathbb L_{+-}\oplus\mathbb L_{-+}\oplus\mathbb L_{--}.
\label{eq:Liouvillian-blocks}
\end{equation}
Each block acts on a $4^{N-1}$-dimensional operator subspace. This decomposition follows solely from the strong $\mathbb Z_2$ symmetry and therefore persists away from self-duality.

The KW defect further selects among these sectors. Since $D=P_+DP_+$, its Liouville-space lift obeys $\mathbb D=\Pi_{++}\mathbb D\Pi_{++}$. Consequently, the defect annihilates the $(+,-)$, $(-,+)$, and $(-,-)$ sectors and acts nontrivially only in the even--even sector.  
Thus all four parity sectors remain part of the Liouvillian dynamics, while the KW defect acts nontrivially only in $\mathbb L_{++}$. Translation symmetry gives $[\mathbb L,\mathbb T]=0$ for arbitrary parameters. At self-duality, $[\mathbb L,\mathbb D]=0$, making $\mathbb L_{++}$ the only parity block with a nontrivial KW symmetry. The explicit block form of $\mathbb D$ and a direct illustration of this projection are given in the End Matter.

\begin{figure}[t!]
	\centering	
 \includegraphics[width=\columnwidth]
  {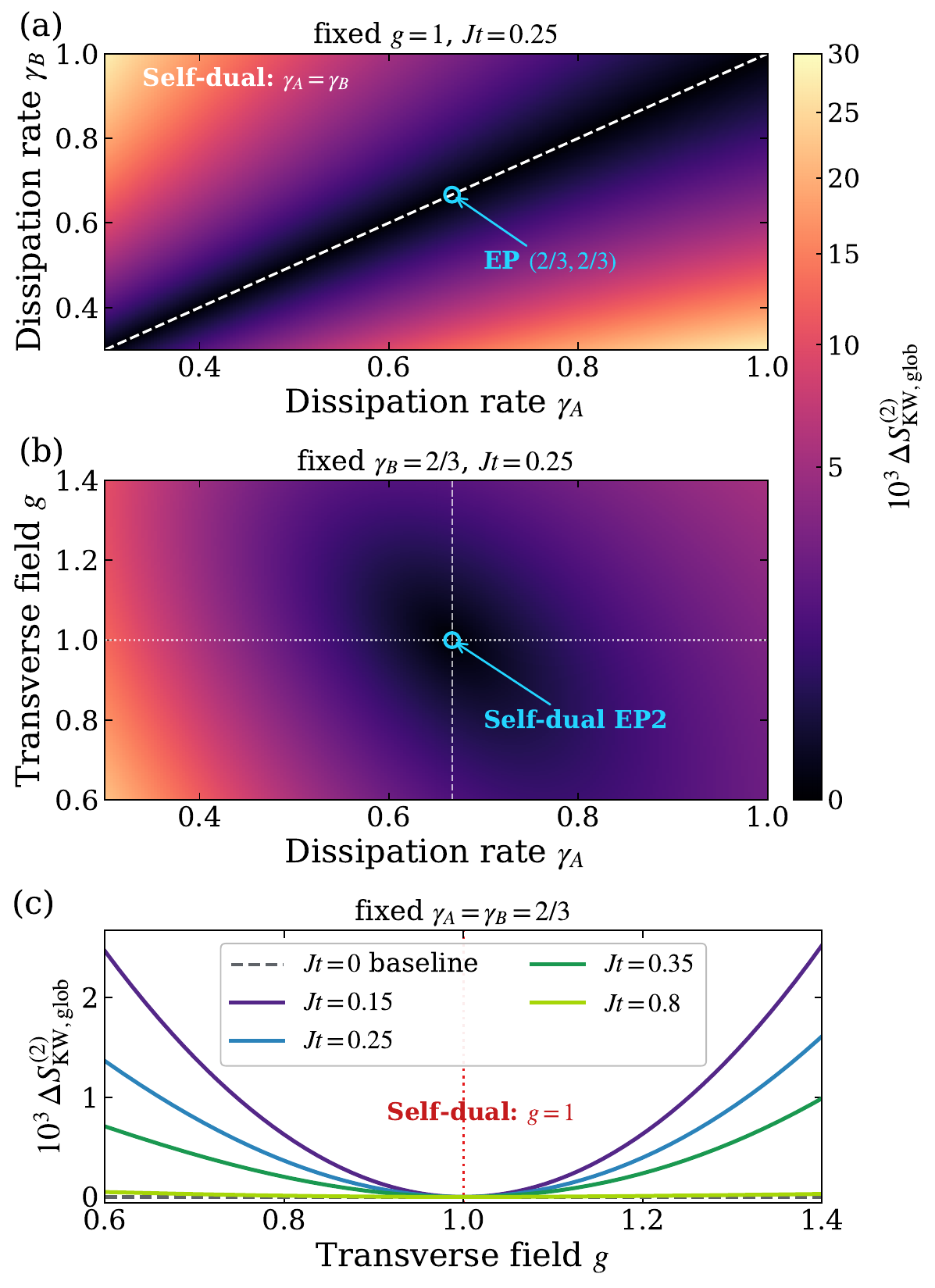} 
\caption{Global second-R\'enyi KW asymmetry for the periodic $N=3$ chain, starting from the fixed state $\rho_0$.  
(a) Rate plane at $g=1$ and $Jt=0.25$. The dashed line denotes $\gamma_A=\gamma_B$, and the cyan circle marks the EP at $(2/3,2/3)$ in $\mathbb L_{++}$.
(b) Field--rate plane at $\gamma_B=2/3$ and $Jt=0.25$. The dotted guides meet at $(\gamma_A,g)=(2/3,1)$.
(c) Field scans at $\gamma_A=\gamma_B=2/3$ for the indicated times.
Panels (a) and (b) use a power-law color normalization with exponent $0.45$. The colorbar gives the untransformed $10^3\Delta S_{\mathrm{KW,glob}}^{(2)}$.}
\label{fig:global-Renyi2-detection}
\end{figure}

\lsec{Dynamical detection of self-duality}
In this section, we investigate whether the self-dual conditions associated with the Liouvillian KW duality relation can be dynamically identified. In Ref.~\cite{Vescovo2026}, the KW entanglement asymmetry was introduced to quantify non-invertible symmetry breaking in the TFIM. Here, we show that the two self-dual conditions, $g=1$ and $\gamma_A=\gamma_B$, can be dynamically identified in a periodic $N=3$ chain. We set $\hbar=1$ and use $J$ as the energy scale. We therefore set $J=1$, with $g$, $\gamma_A$, and $\gamma_B$ measured in units of $J$. Time is expressed as $Jt$. At every point in the parameter scan, we evolve the same initial state,
$|\psi_0\rangle=\frac{\sqrt{3}}{2}|G\rangle+\frac{1}{2}|S\rangle$.
Here, $|G\rangle$ is the GHZ state, and $|S\rangle$ is the normalized equal superposition of the other six computational basis states. The state $|\psi_0\rangle$ is the even-parity, KW-self-dual ground state of the closed $N=3$ TFIM at $g=1$, satisfying $D|\psi_0\rangle=|\psi_0\rangle$. Its density matrix $\rho_0$ is kept fixed throughout the scan, while $g$, $\gamma_A$, and $\gamma_B$ are varied. Following Ref.~\cite{Vescovo2026}, we adapt the KW entanglement asymmetry to the full mixed state, without a spatial partial trace. Defining the dual state $\widehat\rho(t)=D\rho(t)D^\dagger$ and symmetrized state $\rho_{\rm sym}(t)=[\rho(t)+\widehat\rho(t)]/2$, the R\'enyi-2 KW asymmetry is
\begin{equation}
\begin{aligned}
\Delta S_{\mathrm{KW,glob}}^{(2)}(t)
=S^{(2)}\!\left(\rho_{\rm sym}(t)\right)
-S^{(2)}\!\left(\rho(t)\right).
\end{aligned}
\label{eq:global-Renyi2-KW}
\end{equation}
Here $S^{(2)}(\rho)=-\log\operatorname{Tr}\rho^2$.  In the preserved even, translation-invariant sector, $D^2=P_+T$ ensures $\widehat{\widehat\rho}=\rho$.  For the fixed $\rho_0$, the leading dynamical response is
\begin{equation}
\begin{aligned}
\Delta S_{\mathrm{KW,glob}}^{(2)}(t)
&=\ln\frac{4a(t)}{a(t)+b(t)+2c(t)}\\
&=6(Jt)^2\left[(g-1)^2
+(\gamma_A-\gamma_B)^2\right]
+O((Jt)^3),
\end{aligned}
\label{eq:global-Renyi2-short-time}
\end{equation}
where $a(t)=\operatorname{Tr}\rho(t)^2$, $b(t)=\operatorname{Tr}\hat{\rho}(t)^2$, and $c(t)=\operatorname{Tr}\!\left[\rho(t)\hat{\rho}(t)\right]$.
Thus, coherent and dissipative duality breaking contribute independently and quadratically. For $g=1$ and $\gamma_A=\gamma_B$, Liouvillian covariance keeps $\rho(t)=\widehat\rho(t)$ at all times. Breaking either condition produces a finite transient asymmetry. 
We test these signatures in Fig.~\ref{fig:global-Renyi2-detection}.
At $Jt=0$, self-duality of $\rho_0$ makes the signal vanish throughout the scan, so the self-dual conditions cannot yet be identified. At finite time, panel (a) exposes the equal-rate zero line, panel (b) reduces it to a single intersection, and panel (c) resolves the quadratic dip at $g=1$. Relaxation subsequently reduces the contrast, making the transient regime the useful detection window. On the self-dual manifold, the KW-invariant initial state remains
KW invariant under the Liouvillian evolution, so that $\Delta S^{(2)}_{\mathrm{KW,glob}}(t)=0$ for all times, see Eq.~\eqref{eq:global-Renyi2-short-time}. 
The self-dual EP2 therefore lies on the same zero-asymmetry manifold and produces no additional feature in this diagnostic, despite the Jordan-block transient in the underlying Liouvillian dynamics. The R\'enyi asymmetry thus identifies KW self-duality rather than exceptional coalescence, which must instead be resolved from the spectral structure discussed below.

\lsec{Exceptional degeneracies and the KW selection rule}
We now examine how the KW non-invertible symmetry constrains exceptional degeneracies in the Liouvillian. The KW superoperator acts nontrivially only in the even--even sector $\mathcal B_{++}$. Translation symmetry further decomposes the even--even parity sector into Liouville-momentum subspaces, $\mathcal B_{++}=\bigoplus_k\mathcal B_{++,k}$. Within each momentum subspace, the fusion algebra $\mathbb D^2=\Pi_{++}\mathbb T$ implies $\mathbb D_k^2=e^{ik}I$, where $\mathbb D_k$ denotes the restricted KW superoperator. Although $\mathbb D$ is non-invertible on the full Liouville space, $\mathbb D_k$ is invertible on this restricted subspace. Its possible eigenvalues are $d_\pm(k)=\pm e^{ik/2}$.
On the self-dual manifold, $[\mathbb L,\mathbb D]=0$. 
Together with translation symmetry, this implies that the Jordan decomposition of the Liouvillian can be performed separately within subspaces of fixed momentum $k$ and KW-defect eigenvalue $d$. Consequently, modes with different $(k,d)$ cannot belong to the same exceptional Jordan block. Thus, the non-invertible symmetry does not enforce an exceptional point. Instead, it constrains which Liouvillian modes can coalesce into a common exceptional Jordan block. The detailed derivation is given in the End Matter.

We now illustrate this selection rule for the periodic $N=3$ Liouvillian. To distinguish the Hilbert-space momentum from the Liouville-space momentum $k$, we denote the former by $\kappa$. The even-parity Hilbert space contains a two-dimensional $\kappa=0$ subspace spanned by $|G\rangle$ and $|S\rangle$, together with one-dimensional sectors at $\kappa=\pm 2\pi/3$. The corresponding $k=0$ sector of $\mathbb L_{++}$ is six-dimensional. 
Four independent operators span the invariant subspace $\mathcal V_{GS}
=
\operatorname{span}
\left\{
|G\rangle\langle G|,
|S\rangle\langle S|,
|S\rangle\langle G|,
|G\rangle\langle S|
\right\}$.
The remaining two are stationary populations of the $\kappa=\pm2\pi/3$ sectors and decouple from the exceptional structure discussed below. The reduced Liouvillian on $\mathcal V_{GS}$ therefore contains one stationary mode and three nonstationary modes. Their exceptional degeneracies follow from the discriminant of the corresponding cubic characteristic polynomial.

\begin{figure}[t!]
\centering
\includegraphics[width=\columnwidth]{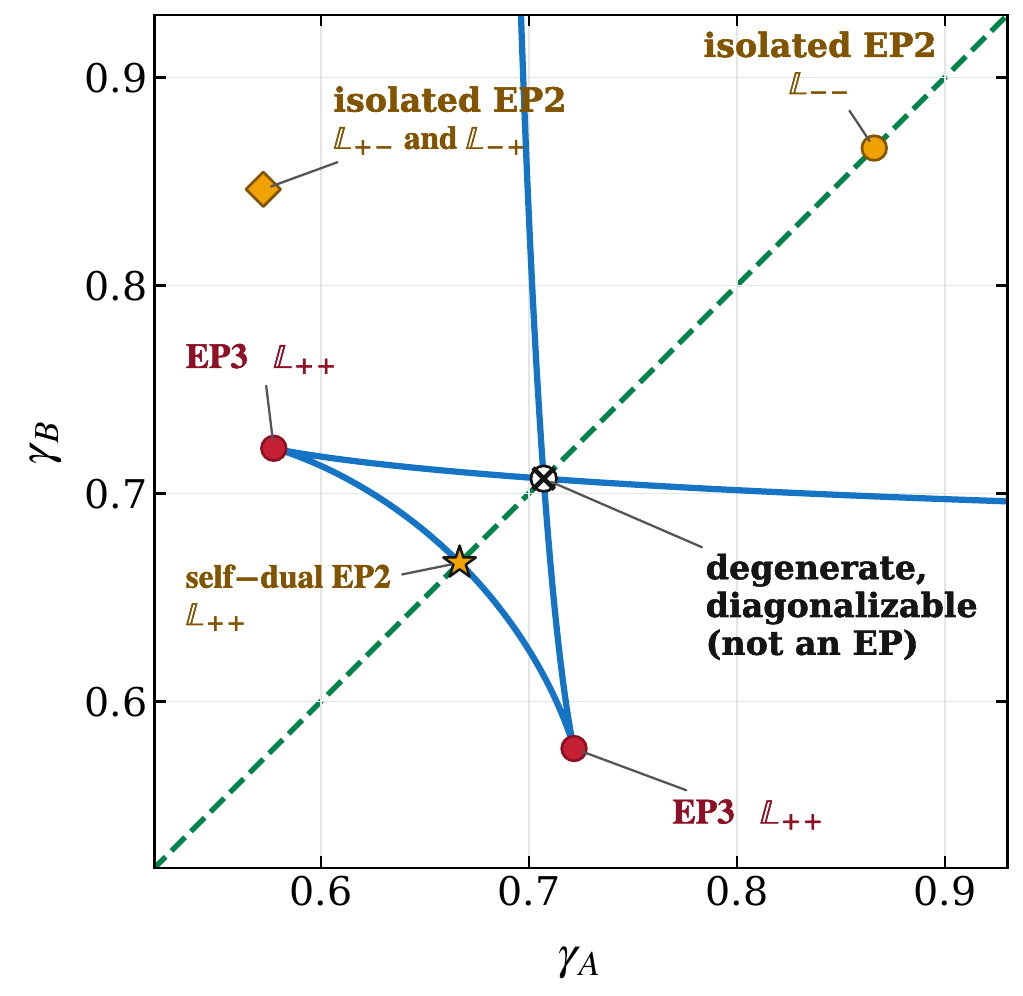}
\caption{Exceptional-degeneracy structure of the periodic $N=3$ Liouvillian at $g=1$. The blue curves denote the zero-discriminant set of the nonstationary block in $\mathbb L_{++}$. Generic points on these curves are EP2s. The red circles mark EP3s, and the black cross marks a semisimple degeneracy. The green dashed line denotes the self-dual condition $\gamma_A=\gamma_B$, whose unique intersection with the EP2 locus in $\mathbb L_{++}$ is marked by the yellow star at $(\gamma_A,\gamma_B)=(2/3,2/3)$. The yellow diamond marks an isolated EP2 shared by $\mathbb L_{+-}$ and $\mathbb L_{-+}$, while the yellow circle marks an isolated EP2 in $\mathbb L_{--}$.}
\label{fig:EP-geometry}
\end{figure}

Figure~\ref{fig:EP-geometry} shows the exceptional-degeneracy structure in the $(\gamma_A,\gamma_B)$ plane at $g=1$. Generic points on the blue branches in $\mathbb L_{++}$ are EP2s, while the two red circles are EP3s. The black cross instead marks a diagonalizable double degeneracy. Since $\mathcal V_{GS}$ lies at Liouville momentum $k=0$, the two KW-defect eigenvalues are $d=\pm1$. At the black cross, $\gamma_A=\gamma_B=1/\sqrt{2}$, the two degenerate modes carry opposite eigenvalues, $d=+1$ and $d=-1$. They therefore remain in distinct invariant subspaces and cannot form a common Jordan block. Since each eigenvalue is simple within its respective subspace, the degeneracy is semisimple.

The self-dual line also intersects the EP2 branch at
\begin{equation}
(g,\gamma_A,\gamma_B)_{\mathrm{sd\text{-}EP2}}
=
\left(1,\frac{2}{3},\frac{2}{3}\right).
\label{eq:self-dual-EP2}
\end{equation}
At this self-dual EP2, both coalescing modes instead have $d=-1$. Their coalescence is therefore allowed by the KW selection rule. The reduced Liouvillian develops a second-order Jordan block at $\lambda_{\mathrm{EP}}=-20/3$, which can produce the characteristic $te^{-20t/3}$ transient for suitable initial conditions.
The explicit $d=\pm1$ decomposition and further spectral details are provided in the Supplemental Material~\cite{SM}.
Figure~\ref{fig:EP-geometry} also shows isolated EP2s in the other parity sectors, where the KW superoperator vanishes. These EPs demonstrate that exceptional points can also occur in other Liouville parity sectors, but the KW defect constrains the Jordan structure of EPs only within the defect-active sector.

\lsec{Perturbation-selective response at the self-dual EP2}
EPs have attracted considerable interest for sensing because a generic perturbation that unfolds an EP of order $n$ can produce a nonanalytic spectral response, $\Delta\lambda\propto|\epsilon|^{1/n}$~\cite{Wiersig2014,Hodaei2017,Chen2017}. Such enhanced eigenvalue splitting, however, does not by itself imply improved measurement precision~\cite{Langbein2018,LauClerk2018,Zhang2019,LoughlinSudhir2024}. In particular, the associated noise can offset the spectral enhancement~\cite{LoughlinSudhir2024}.

At the self-dual EP2, we find a simple perturbation-selective response to the local field $H\to H+\epsilon Z_1$. Because $Z_1$ flips the $\mathbb Z_2$ parity, the induced Liouvillian perturbation has no matrix elements within the $(+,+)$ sector at first order. The reduced EP block supported by $\mathcal V_{GS}$ is therefore not directly perturbed by $Z_1$. Any remaining response must come from coupling to modes outside this sector. In contrast, $H\to H+\epsilon X_1$ preserves parity and has a nonzero projection onto $\mathcal V_{GS}$. It therefore perturbs the reduced EP block directly at first order. The corresponding parity selection rule and its Liouville space derivation are given in the End Matter. This filtering at first order suggests that the self-dual EP2 may provide a natural setting for perturbation-selective EP sensing. However, a full analysis of signal and noise is required to establish a genuine metrological advantage.

\lsec{Conclusion}
In this work, we developed a general Liouville space framework for non-invertible symmetry in Markovian dynamics. We then apply it to the periodic TFIM with two collective Lindblad channels related by KW duality. Exact Liouvillian non-invertible symmetry appears when the coherent and dissipative parameters satisfy the self-dual conditions, $J=g$ and $\gamma_A=\gamma_B$. At self-duality, the KW fusion algebra imposes a momentum-resolved selection rule. Consequently, modes with different $(k,d)$ cannot form a common exceptional Jordan block. This rule constrains exceptional coalescence without enforcing an EP. 
For $N=3$, this rule distinguishes a semisimple degeneracy between modes with opposite KW-defect eigenvalues from a self-dual EP2 formed within the $d=-1$ sector. A global R\'enyi-2 KW asymmetry dynamically identifies the self-dual conditions, while a separate parity selection rule suppresses the direct first-order response of the reduced EP block to a local $Z_1$ perturbation.

The general symmetry construction applies beyond the three-site example, but our explicit spectral analysis and dynamical diagnostic are demonstrated for $N=3$. 
Future work may explore larger systems, other dissipative models, and the duality conditions required beyond the Markovian regime. The interpretation of non-invertible symmetries as quantum operations suggests possible circuit realizations~\cite{OkadaTachikawa2024,AshkenaziZohar2022,Tantivasadakarn2024,Tantivasadakarn2023,LuSunYou2024,Samanta2025,Lamb2025}. Recent studies have connected non-invertible symmetries to robust spectral degeneracies~\cite{CordovaGarciaSepulvedaHolfester2024,LiMitra2026} and conventional symmetries to the stabilization of higher-order EPs~\cite{MandalBergholtz2021}. Non-Markovian quantum EPs provide another direction for investigation~\cite{LinKuoLambert2025,ZhangHanWu2025,WongZengLi2026}. An important open question is whether non-invertible symmetries can enforce or protect exceptional degeneracies, rather than only constrain their allowed Jordan structure.

\lsec{Acknowledgments}
This work is supported by the National Center for Theoretical Sciences and the National Science and Technology Council, Taiwan, under Grant No. NSTC 114-2112-M-006-015-MY3. We thank Christopher Yang for valuable discussions. OpenAI's GPT-6 was used to correct grammatical errors and improve the language and style of the manuscript. All references cited in this work were independently identified and verified by the authors. OpenAI's Astra was used solely to assist with organizing the BibTeX file and standardizing reference formatting.

\lsec{Data availability}
The data that support the findings of this study are available from the first author.

\lsec{Note added}
While completing this work, we became aware of a related study~\cite{teh2026dissipative} that constructs non-invertible Kramers–Wannier symmetries in dissipative Majorana and Ising chains. That work focuses on anomalies and bosonization, whereas ours addresses symmetry constraints on Liouvillian Jordan structures. These studies address complementary aspects of non-invertible symmetries in open quantum systems.

\bibliography{Liouvillian_NIS}

\clearpage

\onecolumngrid
\begin{center}
\ \vskip 0.2cm
{\large\bfseries End Matter}
\end{center}
\twocolumngrid

\renewcommand{\theequation}{A\arabic{equation}}
\stepcounter{superequation}

\phantomsection
\label{sec:EM-KW}
\textit{Direct verification of the two-channel KW duality relation.}---
Here we explicitly derive the Liouvillian KW duality relation used in the main text directly from the vectorized generator.
The Liouville-space KW defect is $\mathbb D=D^*\otimes D$. We denote $H=H(J,g)$ and $H'=H(g,J)$. The lattice KW relations $DA=BD$ and $DB=AD$, obtained by summing the local relations of Ref.~\cite{Shao2023}, imply $DH=H'D$.
We first consider the Hamiltonian contribution.
From $DH=H'D$ and its complex conjugate,
$D^*H^T=H'^T D^*$, one obtains $\mathbb D(I\otimes H)=D^*\otimes DH=D^*\otimes H'D=(I\otimes H')\mathbb D$, whereas $\mathbb D(H^T\otimes I)=D^*H^T\otimes D=H'^T D^*\otimes D=(H'^T\otimes I)\mathbb D$.
Hence
\begin{align}
\mathbb D
\bigl[-i(I\otimes H-H^T\otimes I)\bigr]
=
-i(I\otimes H'-H'^T\otimes I)\mathbb D.
\end{align}
We next consider the dissipative contributions.
We denote by $\mathbb C_O$ the Liouville-space matrix representation
of the dissipative superoperator $\mathcal C_O$ appearing in the
master equation:
\begin{align}
\mathbb C_O
=
O^*\otimes O
-\frac12
\left[
I\otimes O^2+(O^2)^T\otimes I
\right],
\end{align}
where $O=A,B$.
From $DA=BD$, one also has
$DA^2=B^2D$.
Taking the complex conjugate gives
$D^*A^*=B^*D^*$ and
$D^*(A^2)^T=(B^2)^TD^*$.
Therefore,
\[
\begin{aligned}
\mathbb D\mathbb C_A
={}&
D^*A^*\otimes DA
-\frac12D^*\otimes DA^2
-\frac12D^*(A^2)^T\otimes D
\\
={}&
B^*D^*\otimes BD
-\frac12D^*\otimes B^2D
-\frac12(B^2)^TD^*\otimes D
\\
={}&
\left[
B^*\otimes B
-\frac12
\left(
I\otimes B^2+(B^2)^T\otimes I
\right)
\right]
(D^*\otimes D).
\end{aligned}
\]
Thus
\begin{equation}
\mathbb D\mathbb C_A
=
\mathbb C_B\mathbb D.
\label{eq:EM-DA-channel}
\end{equation}
Similarly, $DB=AD$ gives
\begin{equation}
\mathbb D\mathbb C_B
=
\mathbb C_A\mathbb D.
\label{eq:EM-DB-channel}
\end{equation}

We can now act with $\mathbb D$ from the left on the full Liouvillian:
\[
\begin{aligned}
&\mathbb D\,
\mathbb L(J,g;\gamma_A,\gamma_B)
\\
={}&
\mathbb D
\bigl[-i(I\otimes H-H^T\otimes I)\bigr]
+\gamma_A\mathbb D\mathbb C_A
+\gamma_B\mathbb D\mathbb C_B
\\
={}&
-i(I\otimes H'-H'^T\otimes I)\mathbb D
+\gamma_A\mathbb C_B\mathbb D
+\gamma_B\mathbb C_A\mathbb D
\\
={}&
\left[
-i(I\otimes H'-H'^T\otimes I)
+\gamma_B\mathbb C_A
+\gamma_A\mathbb C_B
\right]\mathbb D.
\end{aligned}
\]
The expression in brackets is precisely
$\mathbb L(g,J;\gamma_B,\gamma_A)$.
Therefore, we recover the Liouvillian KW duality relation given in Eq.~\eqref{eq:Liouvillian-covariance}. It exchanges $J\leftrightarrow g$ and $\gamma_A\leftrightarrow\gamma_B$.
Under the self-dual conditions $J=g$ and $\gamma_A=\gamma_B$, the two Liouvillian generators coincide, reducing Eq.~\eqref{eq:Liouvillian-covariance} to $\mathbb D\mathbb L=\mathbb L\mathbb D$. Equivalently, $[\mathbb L,\mathbb D]=0$.
Importantly, this derivation relies only on the KW duality relations of $D$ and does not require an inverse of the non-invertible defect. Thus, the two-channel model realizes the general parameter-space duality described in the main text, with $\sigma(J,g,\gamma_A,\gamma_B)=(g,J,\gamma_B,\gamma_A)$.

\phantomsection
\label{sec:EM-003}
\lsec{Liouville-space action of the KW defect}
We now give an explicit state-level illustration of the defect support discussed in the main text. We then show how this support appears in the four parity sectors of the two-channel Liouvillian. Consider
\begin{equation}
|\psi\rangle
=
a|\psi_+\rangle+b|\psi_-\rangle,
\qquad
|a|^2+|b|^2=1,
\label{eq:EM-psi}
\end{equation}
where $|\psi_\pm\rangle\in\mathcal H_\pm$ and $\eta|\psi_\pm\rangle=\pm|\psi_\pm\rangle$.
The corresponding density matrix is
\begin{align}
\rho
&=|a|^2|\psi_+\rangle\langle\psi_+|
+ab^*|\psi_+\rangle\langle\psi_-| \notag \\
&+a^*b|\psi_-\rangle\langle\psi_+|
+|b|^2|\psi_-\rangle\langle\psi_-|.
\label{eq:EM-rho}
\end{align}

Writing $\rho_{pq}=P_p\rho P_q$, with $P_\pm=(I\pm\eta)/2$, gives $\rho_{++}=|a|^2|\psi_+\rangle\langle\psi_+|$, $\rho_{+-}=ab^*|\psi_+\rangle\langle\psi_-|$, $\rho_{-+}=a^*b|\psi_-\rangle\langle\psi_+|$, and $\rho_{--}=|b|^2|\psi_-\rangle\langle\psi_-|$.
Thus $\rho=\rho_{++}+\rho_{+-}+\rho_{-+}+\rho_{--}$.
The diagonal components contain the weights in the two Hilbert-space parity sectors, whereas the off-diagonal components contain their coherence and relative-phase information.

More generally, the corresponding operator sectors are
$\mathcal B_{pq}=P_p\mathcal B(\mathcal H)P_q$.
Under column vectorization,
$|AOB\rangle\rangle=(B^T\otimes A)|O\rangle\rangle$, so that
\begin{equation}
|\rho_{pq}\rangle\rangle=
\left(P_q^*\otimes P_p\right)|\rho\rangle\rangle.
\label{eq:EM-vectorization}
\end{equation}
The reversed order of $p$ and $q$ in the tensor product follows
directly from this column-vectorization convention.
Defining
$\Pi_{pq}=P_q^*\otimes P_p$, the four components are therefore
selected by
$|\rho_{pq}\rangle\rangle=\Pi_{pq}|\rho\rangle\rangle$.
Because the Hamiltonian and both jump operators commute with $\eta$,
these four sectors are preserved independently by the Liouvillian,
$[\mathbb L,\Pi_{pq}]=0$.

We now apply the KW defect.
Since $D=P_+DP_+$, one has
$DP_p=\delta_{p,+}D$ and
$P_qD^\dagger=\delta_{q,+}D^\dagger$.
For the defect map
$\Phi_D(O)=DOD^\dagger$, this gives
\begin{equation}
\Phi_D(O_{pq})
=\delta_{p,+}\delta_{q,+}\,
D O_{++}D^\dagger.
\label{eq:EM-defect-selection}
\end{equation}

Equivalently,
$\Phi_D(O_{+-})=\Phi_D(O_{-+})=\Phi_D(O_{--})=0$,
while $\Phi_D(O_{++})=DO_{++}D^\dagger$.

For the state in Eq.~\eqref{eq:EM-psi}, all odd-parity weight and
even--odd coherence are therefore removed, leaving
\begin{equation}
\Phi_D(\rho)
=
|a|^2
D|\psi_+\rangle\langle\psi_+|D^\dagger.
\label{eq:EM-example-output}
\end{equation}
The defect annihilates the information carried by $|b|^2$, $ab^*$, and $a^*b$.
This provides a direct illustration of its non-invertibility on the full operator space.

In Liouville space,
$\mathbb D=D^*\otimes D$, and the same selection rule becomes
\begin{equation}
\mathbb D
=
\Pi_{++}\mathbb D\Pi_{++}.
\label{eq:EM-D-support}
\end{equation}
Hence, in the sector ordering $(+,+),(+,-),(-,+)$ and $(-,-)$, the Liouville-space KW defect takes the block form
\begin{equation}
\mathbb D
=
\begin{pmatrix}
\mathbb D_{++} & 0 & 0 & 0 \\
0 & 0 & 0 & 0 \\
0 & 0 & 0 & 0 \\
0 & 0 & 0 & 0
\end{pmatrix}.
\label{eq:EM-D-block}
\end{equation}
Restricted to $\mathcal B_{++}$, the defect map has no kernel. Using the lattice normalization adopted in the main text,
$D^\dagger D=P_+$, and for any
$O_{++}\in\mathcal B_{++}$,
\begin{equation}
D^\dagger\Phi_D(O_{++})D
=
O_{++}.
\label{eq:EM-inverse}
\end{equation}
Thus the non-invertibility of $\mathbb D$ on the full operator space originates from the complete annihilation of the remaining three sectors rather than from its action within $\mathcal B_{++}$. In particular,
\begin{equation}
\ker\mathbb D=
\mathcal B_{+-}
\oplus
\mathcal B_{-+}
\oplus
\mathcal B_{--}.
\label{eq:EM-kernel}
\end{equation}
The four sectors nevertheless remain part of the Liouvillian dynamics, while the KW defect annihilates the three sectors in its kernel. The sector $\mathcal B_{++}$ is therefore distinguished as the only operator sector on which the KW defect acts nontrivially. At self-duality, where $[\mathbb L,\mathbb D]=0$, $\mathbb L_{++}$ is the unique Liouvillian block supporting a nontrivial action of the KW defect.

\textit{KW selection rule for Liouvillian Jordan blocks.—}
Within the defect-active momentum-$k$ sector, $\mathbb D_k^2=e^{ik}I$, so $\mathbb D_k$ is invertible and diagonalizable with eigenvalues $d_\pm(k)=\pm e^{ik/2}$.
At self-duality, $[\mathbb L,\mathbb D]=0$, while translation invariance gives $[\mathbb L,\mathbb T]=0$. The Liouvillian can therefore be restricted to a fixed momentum sector, denoted by $\mathbb L_k$. For an eigenvalue $\lambda$ of $\mathbb L_k$, define the order-$m$ generalized eigenspace by
\begin{equation}
\mathcal G_{\lambda,k}^{(m)}=
\ker\!\left[(\mathbb L_k-\lambda I)^m\right].
\label{eq:EM-generalized-eigenspace}
\end{equation}
Since $[\mathbb L_k,\mathbb D_k]=0$, this generalized eigenspace is invariant under $\mathbb D_k$. Indeed, for $|R\rangle\!\rangle\in\mathcal G_{\lambda,k}^{(m)}$,
\begin{equation}
(\mathbb L_k-\lambda I)^m
\mathbb D_k|R\rangle\!\rangle=0.
\label{eq:EM-generalized-invariance}
\end{equation}
Because $\mathbb D_k$ is diagonalizable, each generalized eigenspace separates into its $d_+(k)$ and $d_-(k)$ eigenspaces. The Jordan decomposition of $\mathbb L_k$ can therefore be performed independently within the two KW-defect eigenspaces. Hence, modes with different $(k,d)$ cannot belong to the same exceptional Jordan block. Equality of $(k,d)$ is only a necessary condition for exceptional coalescence and does not by itself imply the existence of an EP.

\textit{Parity selection rule for the local $Z_1$ perturbation.}---
We now derive the first-order filtering of the local perturbation used in the main text. Consider
\begin{equation}
H\rightarrow H+\epsilon Z_1 ,
\end{equation}
with the corresponding Liouvillian perturbation $\delta\mathbb L_{Z_1} = -i\epsilon \left( I\otimes Z_1-Z_1^T\otimes I \right)$.
The key observation is that $Z_1$ anticommutes with the global spin-flip operator, $\{Z_1,\eta\}=0$, where $\eta=\prod_{j=1}^{N}X_j$.
Hence $Z_1$ maps the even- and odd-parity Hilbert spaces into one another, $Z_1:\mathcal H_{\pm}\rightarrow\mathcal H_{\mp}$, which immediately implies $P_{\pm}Z_1P_{\pm}=0$.

For the $N=3$ invariant subspace relevant to the self-dual EP2, both $|G\rangle$ and $|S\rangle$ belong to $\mathcal H_+$. Defining
\begin{equation}
P_{GS}=|G\rangle\langle G|+|S\rangle\langle S|,
\end{equation}
one therefore obtains $P_{GS}Z_1P_{GS}=0$.
Equivalently, all matrix elements of $Z_1$ within the $\{|G\rangle,|S\rangle\}$ subspace vanish. The projected Hamiltonian is therefore unchanged at first order, $P_{GS}(H+\epsilon Z_1)P_{GS}=P_{GS}HP_{GS}$.

The same selection rule has a direct Liouville-space interpretation. The term $I\otimes Z_1$ flips the ket parity, whereas $Z_1^T\otimes I$ flips the bra parity. Consequently, $\delta\mathbb L_{Z_1}:
\mathcal B_{++}
\longrightarrow
\mathcal B_{-+}\oplus\mathcal B_{+-}$, and hence
\begin{equation}
\Pi_{++}\delta\mathbb L_{Z_1}\Pi_{++}=0.
\label{eq:Pi-Z1-Pi}
\end{equation}
Since the reduced EP subspace $\mathcal V_{GS}$ lies entirely within $\mathcal B_{++}$, with $\mathbb P_{GS}=P_{GS}^{*}\otimes P_{GS}$, Eq.~\eqref{eq:Pi-Z1-Pi} immediately gives
\begin{equation}
\mathbb P_{GS}\delta\mathbb L_{Z_1}\mathbb P_{GS}=0.
\label{eq:PGS-deltaL-PGS}
\end{equation}

Thus the local $Z_1$ perturbation is nonzero in the full Liouville space but has no direct first-order matrix element within the reduced EP block. Any effect on the coalescing modes must therefore arise through coupling to states outside $\mathcal V_{GS}$.

\clearpage
\begin{widetext}
\begin{center}
{\bfseries\Large Supplemental Material for:}\\[0.6em]
{\bfseries\large Liouvillian Non-Invertible Symmetry}\\[0.8em]
\end{center}
\end{widetext}
\vspace{0.8em}

\setcounter{equation}{0}
\setcounter{figure}{0}
\setcounter{table}{0}
\setcounter{section}{0}
\setcounter{subsection}{0}
\makeatletter
\setcounter{secnumdepth}{2}
\renewcommand{\thesection}{\Alph{section}}
\renewcommand{\thesubsection}{\Roman{subsection}}
\renewcommand{\theequation}{S\arabic{equation}}
\renewcommand{\thefigure}{S\arabic{figure}}
\renewcommand{\thetable}{S\arabic{table}}
\renewcommand{\theHfigure}{S\arabic{figure}}
\makeatother

\section{Analytical reduction and KW-resolved exceptional structure in the $(+,+)$ sector}

This Supplemental Material provides an explicit $N=3$ realization of the spectral structure discussed in the main text. We first reduce the periodic Liouvillian to the translation-invariant even--even sector, derive the corresponding $4\times4$ Liouvillian, and obtain its spectrum on the self-dual line. We then resolve this reduced Liouvillian according to the eigenvalues of the KW defect. This decomposition distinguishes the semisimple crossing from the self-dual EP2 and identifies the defect eigenspace in which the exceptional coalescence occurs.

\subsection{Reduced Hilbert- and Liouville-space subspaces}

We set $J=g=1$ and write $H=-(A+B)$, with $A=X_1+X_2+X_3$ and $B=Z_1Z_2+Z_2Z_3+Z_3Z_1$. For the periodic three-site chain, define
\begin{equation}
|G\rangle=\frac{1}{\sqrt{2}}\left(|000\rangle+|111\rangle\right).
\label{eq:SM-G}
\end{equation}
We also define
\begin{equation}
|S\rangle=\frac{1}{\sqrt{6}}\left(|001\rangle+|010\rangle+|100\rangle+|110\rangle+|101\rangle+|011\rangle\right).
\label{eq:SM-S}
\end{equation}
Both states are translation invariant and have even global spin-flip parity, $T|G\rangle=|G\rangle$, $T|S\rangle=|S\rangle$, $\eta|G\rangle=|G\rangle$, and $\eta|S\rangle=|S\rangle$. To distinguish Hilbert-space momentum from the Liouville-space momentum $k$ used in the main text, we denote the former by $\kappa$. The even-parity, zero-momentum Hilbert-space sector is therefore
\begin{equation}
\mathcal H_{\kappa=0,\eta=+1}=\operatorname{span}\{|G\rangle,|S\rangle\}.
\label{eq:SM-HGS}
\end{equation}
The corresponding four-dimensional operator subspace is
\begin{equation}
\mathcal V_{GS}=\operatorname{span}\left\{|G\rangle\langle G|,|S\rangle\langle S|,|S\rangle\langle G|,|G\rangle\langle S|\right\}.
\label{eq:SM-VGS}
\end{equation}
Each basis operator has even parity on both ket and bra indices, so
\begin{equation}
\mathcal V_{GS}\subset\mathcal B_{++}.
\label{eq:SM-VGS-Bpp}
\end{equation}
Since both ket and bra have Hilbert-space momentum $\kappa=0$, every operator in $\mathcal V_{GS}$ has Liouville momentum $k=0$. The full $k=0$ sector of $\mathbb L_{++}$ also contains the populations of the one-dimensional Hilbert-space sectors at $\kappa=\pm2\pi/3$. These two population modes are stationary and decouple from $\mathcal V_{GS}$. The exceptional structure considered below is therefore completely captured by the four-dimensional reduced subspace.

Let
\begin{equation}
U=(|G\rangle,|S\rangle)
\label{eq:SM-U}
\end{equation}
denote the Hilbert-space isometry onto this subspace. Projecting the Ising-bond operator gives
\begin{equation}
B_e=U^\dagger B U=
\begin{pmatrix}
3&0\\
0&-1
\end{pmatrix}.
\label{eq:SM-Be}
\end{equation}
The transverse-field operator becomes
\begin{equation}
A_e=U^\dagger A U=
\begin{pmatrix}
0&\sqrt{3}\\
\sqrt{3}&2
\end{pmatrix}.
\label{eq:SM-Ae}
\end{equation}
The reduced Hamiltonian is therefore
\begin{equation}
H_e=-U^\dagger(A+B)U=
\begin{pmatrix}
-3&-\sqrt{3}\\
-\sqrt{3}&-1
\end{pmatrix}.
\label{eq:SM-He}
\end{equation}

\subsection{Reduced $4\times4$ Liouvillian}

For the Hermitian collective operators $A$ and $B$, the reduced dynamics follows from
\begin{equation}
\mathcal L(\rho)=-i[H,\rho]+\gamma_A\mathcal C_A(\rho)+\gamma_B\mathcal C_B(\rho).
\label{eq:SM-master}
\end{equation}
For a Hermitian operator $O$, the dissipator is
\begin{equation}
\mathcal C_O(\rho)=O\rho O-\frac{1}{2}\{O^2,\rho\}.
\label{eq:SM-dissipator}
\end{equation}
We order the operator basis of $\mathcal V_{GS}$ as $E_1=|G\rangle\langle G|$, $E_2=|S\rangle\langle S|$, $E_3=|S\rangle\langle G|$, and $E_4=|G\rangle\langle S|$. The reduced vectorized density matrix is
\begin{equation}
|\rho\rangle\!\rangle_{\mathrm{red}}=(\rho_{GG},\rho_{SS},\rho_{SG},\rho_{GS})^T.
\label{eq:SM-rho-red}
\end{equation}
Its equation of motion is
\begin{equation}
\partial_t|\rho\rangle\!\rangle_{\mathrm{red}}=M_4(\gamma_A,\gamma_B)|\rho\rangle\!\rangle_{\mathrm{red}}.
\label{eq:SM-eom-red}
\end{equation}
The reduced Liouvillian matrix is
\begin{widetext}
\begin{equation}
M_4(\gamma_A,\gamma_B)=
\begin{pmatrix}
-3\gamma_A & 3\gamma_A & \sqrt3(-\gamma_A+i) & \sqrt3(-\gamma_A-i)\\
3\gamma_A & -3\gamma_A & \sqrt3(\gamma_A-i) & \sqrt3(\gamma_A+i)\\
\sqrt3(i-\gamma_A) & \sqrt3(\gamma_A-i) & -5\gamma_A-8\gamma_B-2i & 3\gamma_A\\
\sqrt3(-i-\gamma_A) & \sqrt3(\gamma_A+i) & 3\gamma_A & -5\gamma_A-8\gamma_B+2i
\end{pmatrix}.
\label{eq:SM-M4}
\end{equation}
\end{widetext}

\subsection{Self-dual spectrum and KW-defect eigenspaces}

On the self-dual line we set
\begin{equation}
\gamma_A=\gamma_B\equiv\gamma.
\label{eq:SM-selfdual-gamma}
\end{equation}
The characteristic polynomial of the reduced Liouvillian factorizes as
\begin{equation}
\det(\lambda I_4-M_4)=\lambda(\lambda+12\gamma)\left(\lambda^2+20\gamma\lambda+64\gamma^2+16\right).
\label{eq:SM-charpoly}
\end{equation}
One eigenvalue is
\begin{equation}
\lambda_0=0.
\label{eq:SM-lambda0}
\end{equation}
A second eigenvalue is
\begin{equation}
\lambda_1=-12\gamma.
\label{eq:SM-lambda1}
\end{equation}
The remaining two eigenvalues are
\begin{equation}
\lambda_\pm=-10\gamma\pm2\sqrt{9\gamma^2-4}.
\label{eq:SM-lambdapm}
\end{equation}

We now resolve these four branches according to the KW defect. Restricting the Hilbert-space KW operator to $\operatorname{span}\{|G\rangle,|S\rangle\}$ gives
\begin{equation}
D_e=U^\dagger D U=\frac{1}{2}
\begin{pmatrix}
1&\sqrt3\\
\sqrt3&-1
\end{pmatrix}.
\label{eq:SM-De}
\end{equation}
At zero momentum this restricted defect satisfies
\begin{equation}
D_e^2=I_2.
\label{eq:SM-De-square}
\end{equation}
Its eigenstate with eigenvalue $+1$ is
\begin{equation}
|\phi_+\rangle=\frac{\sqrt3}{2}|G\rangle+\frac{1}{2}|S\rangle.
\label{eq:SM-phiplus}
\end{equation}
Its eigenstate with eigenvalue $-1$ is
\begin{equation}
|\phi_-\rangle=-\frac{1}{2}|G\rangle+\frac{\sqrt3}{2}|S\rangle.
\label{eq:SM-phiminus}
\end{equation}
The Liouville-space defect restricted to $\mathcal V_{GS}$ is $\mathbb D_{GS}=D_e^*\otimes D_e$. We choose the symmetry-adapted operator basis $F_1=|\phi_+\rangle\langle\phi_+|$, $F_2=|\phi_-\rangle\langle\phi_-|$, $F_3=|\phi_+\rangle\langle\phi_-|$, and $F_4=|\phi_-\rangle\langle\phi_+|$. In this basis,
\begin{equation}
\mathbb D_{GS}=\operatorname{diag}(1,1,-1,-1).
\label{eq:SM-DGS}
\end{equation}
Thus $F_1$ and $F_2$ span the $d=+1$ eigenspace, whereas $F_3$ and $F_4$ span the $d=-1$ eigenspace. On the self-dual line, $[M_4,\mathbb D_{GS}]=0$, so the reduced Liouvillian becomes block diagonal in the same basis,
\begin{equation}
\widetilde M_4=M_+\oplus M_-.
\label{eq:SM-M4-block}
\end{equation}
The $d=+1$ block is
\begin{equation}
M_+(\gamma)=
\begin{pmatrix}
-6\gamma&6\gamma\\
6\gamma&-6\gamma
\end{pmatrix}.
\label{eq:SM-Mplus}
\end{equation}
The $d=-1$ block is
\begin{equation}
M_-(\gamma)=
\begin{pmatrix}
-10\gamma+4i&6\gamma\\
6\gamma&-10\gamma-4i
\end{pmatrix}.
\label{eq:SM-Mminus}
\end{equation}
The spectrum of $M_+$ is $\lambda_0$ and $\lambda_1$, whereas the spectrum of $M_-$ is $\lambda_+$ and $\lambda_-$. The stationary branch and the $-12\gamma$ branch belong to the $d=+1$ eigenspace. The pair that produces the self-dual EP2 belongs entirely to the $d=-1$ eigenspace.

\subsection{Semisimple crossing and the self-dual EP2}

The KW-resolved form distinguishes two qualitatively different degeneracies on the self-dual line. First consider $\gamma=1/\sqrt2$. At this point,
\begin{equation}
\lambda_1=\lambda_-=-6\sqrt2.
\label{eq:SM-crossing-eigenvalue}
\end{equation}
The branch $\lambda_1$ belongs to the $d=+1$ block, whereas $\lambda_-$ belongs to the $d=-1$ block. The two degenerate modes therefore remain in distinct invariant eigenspaces of the KW defect and cannot form a common Jordan block. Since the corresponding eigenvalue is simple within each $2\times2$ block, the full degeneracy has two independent eigenvectors and is semisimple. This is the black crossing shown in Fig.~3 of the main text.

The situation is different at $\gamma=2/3$. The two eigenvalues within the same $d=-1$ block coalesce,
\begin{equation}
\lambda_+=\lambda_-=-\frac{20}{3}.
\label{eq:SM-EP-eigenvalue}
\end{equation}
At this point the $d=-1$ block becomes
\begin{equation}
M_-\!\left(\frac{2}{3}\right)=
\begin{pmatrix}
-\frac{20}{3}+4i&4\\
4&-\frac{20}{3}-4i
\end{pmatrix}.
\label{eq:SM-Mminus-EP}
\end{equation}
Subtracting the coalesced eigenvalue defines
\begin{equation}
N_-=M_-\!\left(\frac{2}{3}\right)+\frac{20}{3}I_2.
\label{eq:SM-Nminus-def}
\end{equation}
Explicitly,
\begin{equation}
N_-=4
\begin{pmatrix}
i&1\\
1&-i
\end{pmatrix}.
\label{eq:SM-Nminus}
\end{equation}
Its kernel is one-dimensional,
\begin{equation}
\dim\ker N_-=1.
\label{eq:SM-Nminus-nullity}
\end{equation}
The shifted block is nilpotent of order two,
\begin{equation}
N_-^2=0.
\label{eq:SM-Nminus-square}
\end{equation}
The coalesced eigenvalue therefore has algebraic multiplicity two but geometric multiplicity one, establishing a genuine EP2 entirely within the $d=-1$ eigenspace. The $d=+1$ block remains diagonalizable, with eigenvalues $0$ and $-8$, and does not participate in the exceptional coalescence. This gives an explicit $N=3$ realization of the KW selection rule. Opposite values of $d$ forbid a common Jordan block at the semisimple crossing. Equal $d$ allows the EP2 but does not by itself enforce an exceptional degeneracy.

\subsection{Jordan evolution at the self-dual EP2}

The defective $2\times2$ block is similar to a Jordan block of the form
\begin{equation}
J_{\mathrm{EP}}=\lambda_{\mathrm{EP}}I_2+N_J.
\label{eq:SM-JEP}
\end{equation}
Its nilpotent part is
\begin{equation}
N_J=
\begin{pmatrix}
0&1\\
0&0
\end{pmatrix}.
\label{eq:SM-NJ}
\end{equation}
The nilpotent part satisfies
\begin{equation}
N_J^2=0.
\label{eq:SM-NJ-square}
\end{equation}
The propagator therefore factorizes as
\begin{equation}
e^{J_{\mathrm{EP}}t}=e^{\lambda_{\mathrm{EP}}t}e^{N_Jt}.
\label{eq:SM-Jprop-factor}
\end{equation}
Because $N_J^2=0$, its exponential truncates exactly,
\begin{equation}
e^{N_Jt}=I_2+tN_J.
\label{eq:SM-NJ-exp}
\end{equation}
Using $\lambda_{\mathrm{EP}}=-20/3$ gives
\begin{equation}
e^{J_{\mathrm{EP}}t}=e^{-20t/3}
\begin{pmatrix}
1&t\\
0&1
\end{pmatrix}.
\label{eq:SM-Jprop}
\end{equation}
An initial condition with a component along the generalized eigenvector therefore produces the characteristic transient
\begin{equation}
t e^{-20t/3}.
\label{eq:SM-EP-transient}
\end{equation}
The reduced $N=3$ construction thus gives a direct analytic connection between the KW-defect eigenspaces and the exceptional structure of the Liouvillian. The semisimple crossing occurs between opposite defect eigenvalues, whereas the self-dual EP2 is confined to the $d=-1$ eigenspace and produces the characteristic nonexponential transient.

\end{document}